\documentclass[review]{elsarticle}

\usepackage{hyperref}

\journal{Computer Physics Communication}

\usepackage{amssymb}  
\usepackage{amsthm}
\usepackage{amsmath}

\usepackage[numbers]{natbib}

\usepackage{tikz}

\def\<{\langle}
\def\>{\rangle}

\newcommand{\numberedgraphics}[5]{
	\begin{tikzpicture}
	\draw node[name=micrograph] {\includegraphics[width=#2\textwidth]{#1}}; 
	\draw  (micrograph.north west)  node[yshift=#4*#2*\textwidth, xshift=#3*#2*\textwidth]{\small{(#5)}}; 
	\end{tikzpicture}}

\begin{document}

	\begin{frontmatter}
		
		\title{On the Delta Function Broadening in Kubo-Greenwood Equation}
		
		\author{Pavlo Bulanchuk}
		\address{Pennsylvania State University, 104 Davey lab, University Park, Pennsylvania}
		
		\begin{abstract}
			Understanding DC electrical conductivity is crucial for the study of materials. Macroscopic DC conductivity can be calculated from first principles using the Kubo-Greenwood equation. The procedure involves finding the thermodynamic limit of the current response to an electric field that is slowly switched on, and then taking the limit of the switching rate to zero. 
			We introduce a nonlinear extrapolation procedure executed in systems with periodic boundary conditions, which predicts conductivity close to the thermodynamic limit even for very small systems.
			The scheme also overcomes a large part of the usual ambiguities of the DC conductivity definition for finite systems. We numerically compare our method to the Landauer equation-based approach and find both techniques to be consistent with each other.  
		\end{abstract}
		
		
	\end{frontmatter}

	\section{Introduction}
	
	This work was motivated by the issue of ambiguity of the definition of DC conductivity in finite-sized quantum systems, calculated by Kubo's linear response theory \cite{cald1}. The Kubo-Greenwood formula  \cite{kubo1,gree1} expresses the real part of AC electrical conductivity $\sigma(\omega)$ as a sum of delta functions:
	\begin{equation}
	\mathrm{Re}[\sigma(\omega)]=\sum_{mn}\Gamma_{mn} \delta(E_m-E_n-\omega),
	\label{kubo}
	\end{equation}
	where $E_n$ are eigenstate energies, $\omega$ is the frequency of the external electric field, and $\Gamma_{mn}$ depends on the Hamiltonian and temperature of the system, but not on $\omega$. Multiple equivalent forms of Eq.~\eqref{kubo} can be found in the literature \cite{cald1, gree1,desj1, knya1, nomu1, arab1, czyc1, thou1, silv1, silv2}. Conductivity $\sigma(\omega)$ is defined as a Fourier transform of the time dependent $\sigma(t)$, which a part of the linear response equation for current density $j(t)$:
	\begin{equation}
	j(t)=\int_{-\infty}^t \sigma(t-t') E(t')dt'+O(E^2),\,
	\label{linearResponse00}
	\end{equation}
	where $E(t)$ is a time dependent electric field.
	
	Due to the delta function in Eq.~\eqref{kubo}, DC conductivity is always zero for closed and finite systems. A non-zero DC conductivity can be defined rigorously as a result of taking two consecutive limits, one of which includes taking the size of the system to infinity:
	\begin{enumerate}
		\item First, one defines a \textit{pseudo}-conductivity $\text{Re}[\sigma(\omega)]_{\eta}$, where each delta function in Eq.~(\ref{kubo}) is exchanged for a smooth representation $\delta(\omega)\rightarrow\delta_{\eta}(\omega)$, where $\eta$ is an effective width of the representation. For example $\delta_{\eta}(\omega)=\eta^{-1}\pi^{-1/2}e^{-\omega^2/\eta^2}$ is a valid smearing function. 
		\item A finite temperature DC conductivity is defined as a result of two consecutive limits:
		\begin{equation}
		\sigma_{DC}\equiv\lim_{\eta\to 0}\left(\lim_{L\to \infty} \text{Re}[\sigma(0)]_{\eta}\right),
		\label{defDC}
		\end{equation}
		where $L$ is the linear size of system.
	\end{enumerate}
	
	In experiments, DC conductivity of finite systems is not zero, which pushes us to modify Eq.~\eqref{kubo}. Theoretically, finite-sized DC conductivity can be defined by employing additional constructs like infinite electric leads or a thermostat \cite{imry1}. Although these definitions allow us to talk about DC conductivity of a finite system, each of them involves some freedom and can yield different results, i.e., there is no unambiguous definition of DC conductivity for finite systems.
	
	In this paper, we are primarily concerned with the definitions based on the artificial broadening of the delta function in Eq.~(\ref{kubo}) \cite{cald1, gree1,desj1, knya1, nomu1, arab1, czyc1, thou1, abte1}. The methods are usually reasoned empirically: a physically meaningful DC conductivity can be obtained from the Kubo-Greenwood equation only if the delta functions are broadened. If the broadening is small, $\eta\ll\Delta E$ ($\Delta E$ is the distance between energy levels), conductivity $\text{Re}[\sigma(\omega)]_{\eta}$ will experience strong oscillations due to the discrete energy spectrum. If the broadening is large, the features of conductivity, like resonant peaks, may get washed out. Therefore, one hopes to have a region of the broadening $\eta$, where the oscillations from the discrete spectrum are suppressed, but macroscopic features are not washed out. DC conductivity then can be defined as a value $\text{Re}[\sigma(0)]_{\eta}$ obtained from this region. Unfortunately, as Calderin et al. demonstrated in \cite{cald1}, such an approach may yield significantly different values of DC conductivity for small systems depending on the form of the broadening function $\delta_\eta(\omega)$ and the chosen value of $\eta$. The problem is more pronounced for systems with a sharp conductivity peak at $\omega=0$. The peak gets quickly suppressed if the delta function is broadened but has a width comparable to the distance between energy levels. Obtaining DC conductivity for such systems may be difficult and requires consideration of very large systems. 
	
\begin{figure}
		\numberedgraphics{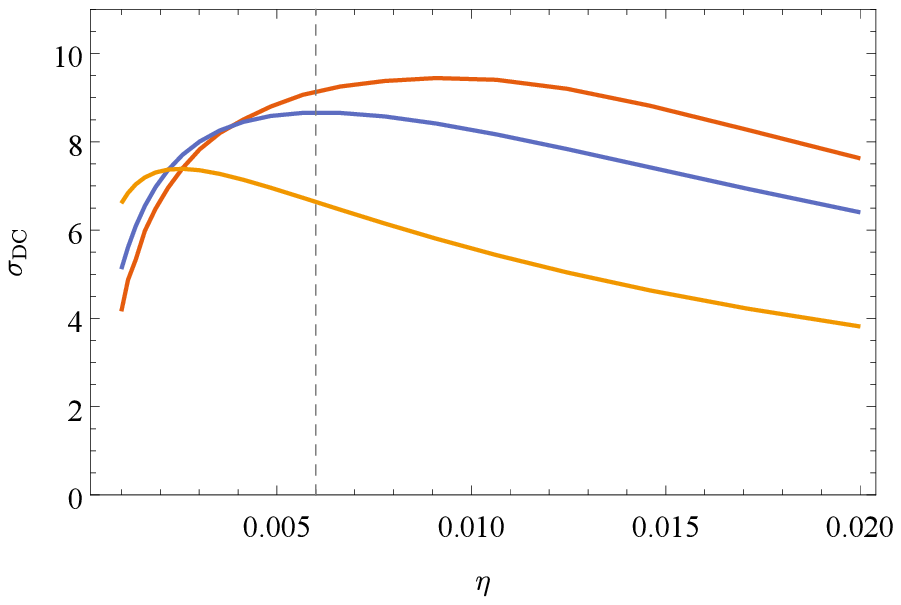}{0.45}{0.15}{-0.06}{a}\numberedgraphics{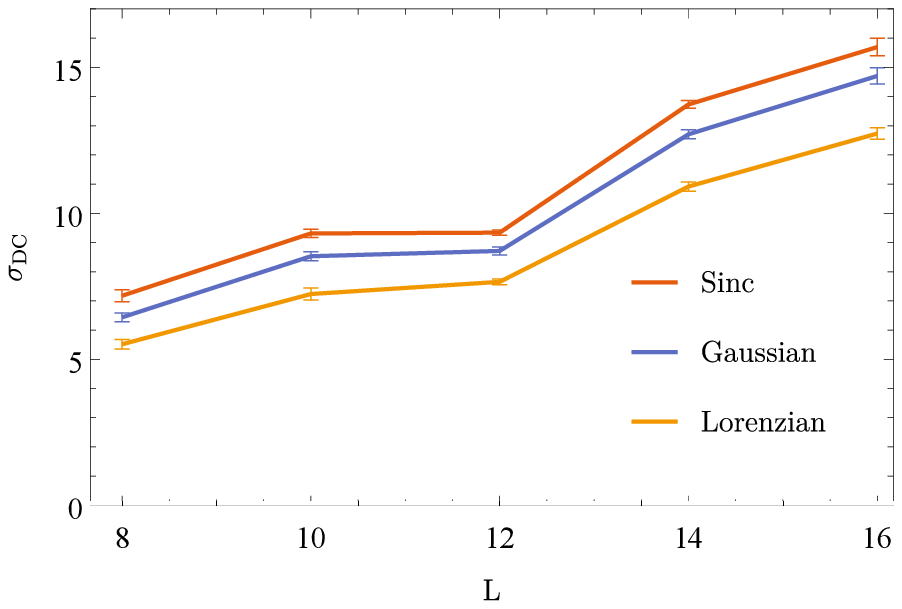}{0.46}{0.16}{-0.06}{b}
		\caption{a) DC conductivity by Eq.~(\ref{kubo_delta_1}) as a function of the quasi-delta-function width $\eta$ for different types of broadening. The red curve corresponds to Lorenzian  broadening  $\delta_\eta(\omega)= \pi^{-1}\eta/(\eta^2+\omega^2)$, blue - to Gaussian $\delta(\omega)= \pi^{-1/2}\exp(-\omega^2/\eta^2)/\eta$, orange - to a sinc broadening $\delta_\eta(\omega)= \pi^{-1}\sin(\omega/2\eta)/\omega$. The system considered is a 3D tight-binding model of spinless, non-interacting fermions, $10\times 10\times 10$ cubic lattice, periodic boundary condition, chemical potential $\mu=-4$, temperature $T=0.6$, inter-cite hopping $t=-1$ and  on-cite energy distributed uniformly in the range $[-\Delta\varepsilon,\Delta\varepsilon]$, where $\Delta\varepsilon=0.1$. Averaging over 16 systems was performed. The dashed line corresponds to the mean inter-level energy difference. b) DC conductivity vs size of the system $L$ for different types of the delta function broadening. Here the DC conductivity is defined as a maximum of $\text{Re}[\sigma_{\text{DC}}(\omega=0)]_{\eta}$ as a function of $\eta$. The graph illustrates extremely slow convergence of the predicted conductivity as the size of the system increases}
		\label{fig:init}
	\end{figure}

	As an illustration of the scope of the issue, we plotted DC conductivity calculated by equation \eqref{kubo} with different types of smearing for a tight-binding nearest-neighbor hopping 3D system with a small onsite disorder. Conductivity as a function of the broadening parameter $\eta$ is presented on Fig. \ref{fig:init}a. From the picture, apart from the arguable absence of the  \textit{plateau} for each curve, there is another problem: the peak heights for different types of broadening differ up to 30\%, and do not converge as the system size increases (see Fig. \ref{fig:init}b). Considering larger systems quickly becomes computationally intractable without radical optimisations (see Kernel Polynomial Method \cite{alex1}), since the eigenvalue problem scales as $O(L^9)$. Extracting DC conductivity by the basic method from small systems seems impossible.

	In this paper we provide a solution to the aforementioned problem which allows extraction of macroscopic DC conductivity even from tiny systems.
	Instead of a peak height, we propose to define a finite-sized DC conductivity based on a specific non-linear extrapolation of $\text{Re}[\sigma(0)]_{\eta}$ from finite $\eta$ to zero. The scheme is based on the equivalence between delta function broadening and current response to variable electric field: $\text{Re}[\sigma(0)]_{\eta}$ is a current response to electric field $E(\omega)=\delta_{\eta}(\omega)$ at $t=0$.  If $\text{Re}[\sigma(0)]_{\eta}$ has reached the thermodynamic limit and macroscopic conductivity follows the Drude equation
	\begin{equation}
	{\text{Re}}[\sigma(\omega)]=A/(1+B^2 \omega^2),
	\label{DrudeIntro}
	\end{equation}
	where $A$ and $B$ are constants, we can predict $\text{Re}[\sigma(0)]_{\eta}$ by calculating response to the corresponding electric field and then obtain DC conductivity $\sigma_{\text{DC}}=A$ by extrapolating from finite $\eta$ to zero. We call the procedure a Drude-based extrapolation because of the assumption of macroscopic conductivity to follow Drude equation.
	
	It turns out if the temperature is high enough, the result of the extrapolation is practically independent of the choice $\delta_\eta(\omega)$, and the value of finite-size DC conductivity converges rapidly to the thermodynamic limit, i.e., we get close to the limit even for very small systems. Conductivity obtained by such extrapolation also matches well with values obtained from the Laundauer-based approach \cite{imry1}, where the system is attached to infinite electric leads, and DC conductivity is extracted from the change of resistance, as the size of the system increases. In summary, for finite system sizes our approach overcomes a large part of the usual ambiguities and provides a reasonable approximation of the macroscopic DC conductivity.\\ 
	
	The paper is structured as follows: We start with a quick review of the Kubo-Greenwood equation and various associated definitions in Section~\ref{sec:summary-Kubo-Greenwood}. We then proceed to the description of the method and numerical tests in Section \ref{sec:method-description}. Then we apply the method to two other toy models: one is a 3D system with point defects, and the other is a 1D disordered chain. 
	
	\section{Summary of the Kubo-Greenwood equation}\label{sec:summary-Kubo-Greenwood}
	In this section, we quickly summarise the general notions of linear response and the Kubo-Greenwood equation. 
	
	\subsection{General results from linear response}
	Linear response of the current of the system to an electric field can be expressed as follows:
	\begin{equation}
	j(t)=\int_{-\infty}^t \sigma(t-t') E(t')dt'\,,
	\label{linearResponse0}
	\end{equation}
	where $E(t)$ is external electric field and $\sigma(t)$ is conductivity. Generally $\sigma(t)$ is a tensor, but we consider only an isotropic case for simplicity. For the same reason, we also assume electric field and current to be directed along the $x$ axis. Without loss of generality we can also choose $t=0$ and assume $E(t')=E(-t')$, since due to causality, fields at $t>0$ should not influence current at $t=0$. Under the above condition, it is possible to express current at $t=0$ in Eq.~(\ref{linearResponse0}) using only real parts of $\sigma$ and $E$ in Fourier space
	\begin{equation}
	j(0)=2\pi \int_{-\infty}^{\infty} \text{Re}[\sigma(\omega)] E(\omega)d\omega\,,
	\label{linearResponse1}
	\end{equation}
	where $\sigma(\omega)=\frac{1}{2\pi}\int \sigma(t) e^{-i\omega t} dt$. Notice how Eq.~(\ref{linearResponse1}) provides a direct way of expressing current using only real part of conductivity.
	
	\subsection{Kubo-Greenwood equation and associate definitions summary}
	In all equations we assume $\hbar=e=1$, where $\hbar$ is Plank constant and $e$ is electron charge. 
	
	Kubo-Greenwood equation predicts $\text{Re}[\sigma(\omega)]$ for a quantum system with finite number of energy levels (see Appendix A for derivation):
	\begin{equation}
	\text{Re}\left[\sigma(\omega)\right]=-\frac{\pi}{Z} \sum_{nm} \frac{e^{-\beta E_n}-e^{-\beta E_n}}{E_n-E_m}j_{nm}j_{mn}\delta(E_n-E_m-\omega),
	\label{kubo_delta0}
	\end{equation}
	where $E_n$ are energies of the system, $Z\equiv\sum_n e^{-\beta E_n}$ is the partition function, $j_{nm}\equiv\<n|\hat{j_x}|m\> $ are matrix elements of the current density operator $\hat{j}_x=-V^{-1}\partial{\hat{H}}/\partial{A_x}$, where $A_x$ is a vector potential along the x-axis, $\hat{H}$ is the Hamiltonian and $V$ is the volume of the system. The dimensionality of the system is arbitrary. 
	
	If the system consists of non-interacting particles, equation (\ref{kubo_delta0}) reduces to the following form
	\begin{equation}
	\text{Re}\left[\sigma(\omega)\right]=-\pi \sum_{nm} \frac{f_n-f_m}{\varepsilon_n-\varepsilon_m}j_{nm}j_{mn}\delta(\varepsilon_n-\varepsilon_m-\omega)\,,
	\label{kubo_delta_1}
	\end{equation}
	where $\varepsilon_n$ are one-electron energies, and $f_n=(1+e^{\beta (\varepsilon_n-\mu)})^{-1}$ is the Fermi distribution.

	Connecting this equation to the Eq.~(\ref{kubo}) in the introduction, we obtain 
	\begin{equation}
	\Gamma_{mn}=-\pi \frac{f_n-f_m}{\varepsilon_m-\varepsilon_n}j_{nm}j_{mn}\,.
	\end{equation}
	Substituting equation for conductivity (\ref{kubo_delta_1}) into (\ref{linearResponse1}) yields the prediction for the current at $t=0$
	
	\begin{equation}
	j(0)=  \sum_{mn}\Gamma_{mn} 2\pi E(\varepsilon_n-\varepsilon_m-\omega)|_{\omega=0}\,.
	\label{linearResponse2}
	\end{equation}
	The form of this equation coincides with the equation for DC conductivity (\ref{kubo}), where each delta function is substituted by $\delta(\omega)\rightarrow 2\pi E(\omega)$. Smearing delta function by a \textit{narrow} function $\delta(\omega)_{\eta}$ corresponds to a \textit{wide} function $E(t)$ in time domain, or, in other words, slowly switched electric field. 
	
	\begin{figure*}
		\numberedgraphics{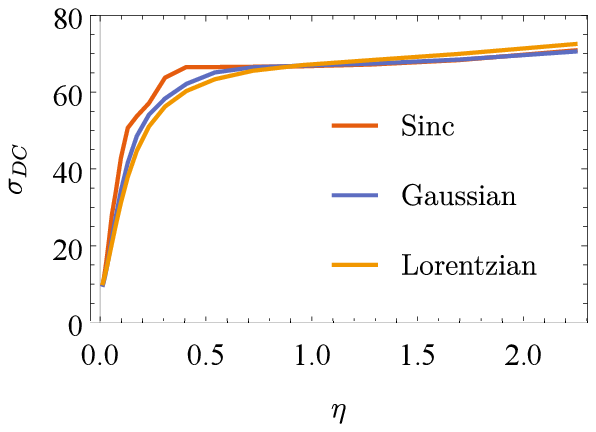}{0.3}{0.22}{-0.1}{a}
		\numberedgraphics{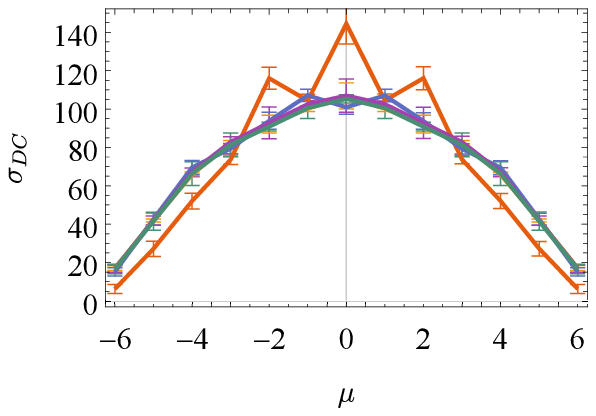}{0.3}{0.25}{-0.1}{b}
		\numberedgraphics{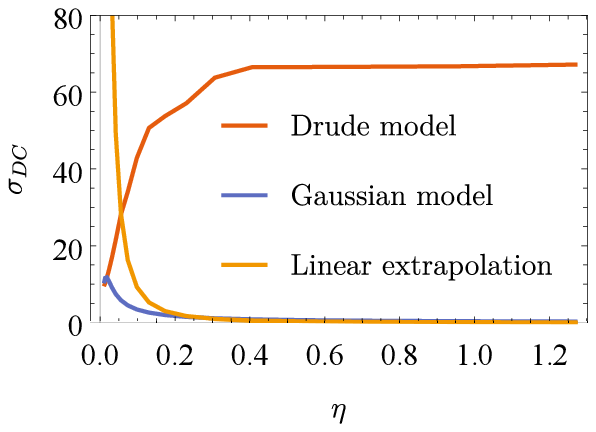}{0.3}{0.28}{-0.1}{c}
		
		\caption{a) Drude-based extrapolated DC conductivity as a function of the fitting region for Lorentzian (orange line), Gausian (blue line) and Sinc (red line) smearing functions. Extrapolation was performed based on the values of quasi-conductivity $\sigma_{\text{DC}\eta'}$ at two points $\eta'=\eta$ and $\eta'=1.25\eta$. The description of the system can be found on Fig. 1. Averaging was performed over 256 samples. b) Predicted DC conductivity as function of the chemical potential $\mu$ for different sizes of the system $L\times L\times L$: L=4 (red curve), L=6 (blue curve), L=8,10,12 (other colors). Averaging is performed over 64 sample systems. c) Predicted DC conductivity for Sinc broadening assuming different models for conductivity. The red curve corresponds to the Drude model, the blue - to a Gaussian model, the orange - to linear extrapolation for comparison. }
		\label{fig:method}
	\end{figure*}
	
	In thermodynamic limit, one usually expects a constant current in response to the constant electric field while in finite systems the current starts dropping after the time $t\gtrsim 1/\Delta E$, where $\Delta E$ is an energy difference between nearby energy levels. This can be seen from the form of equation (\ref{linearResponse2}). A long-lasting electric field corresponds to a narrow delta function. Due to the repulsion of the energy levels, probability density that a delta function is located at $\omega=0$ is zero, so the DC conductivity will tend to zero, as the width of delta function diminishes.

	\section{Formulation of the method}
	\label{sec:method-description}
	
	This section explains the motivation behind our method and demonstrates its numerical advantage over the classical approach. 
	
	\subsection{Drude based extrapolation}
	Our prime interest was macroscopic DC conductivity obtained from the Eq.~\eqref{defDC}. Suppose we found a numerical estimate for thermodynamic limit
	\begin{equation}
	\lim_{L\to \infty}{\text{Re}}[\sigma(0)]_{\eta}\approx{\text{Re}}[\sigma(0)]_{\text{Num}\,\eta}
	\end{equation} 
	for some values of $\eta$.
	In order to find the second limit $\eta\to 0$, let us assume that the ``true'' conductivity of a macroscopic system follows the Drude equation for small $\omega$:
	\begin{equation}
	{\text{Re}}[\sigma(\omega)]_{\text{Drude}}=A/(1+B^2 \omega^2)\,.
	\label{Drude}
	\end{equation}
	Remembering the duality between variable electric field and broadening of the delta function, and using Eq.~(\ref{linearResponse1}), the Drude equation (\ref{Drude}) predicts conductivity for arbitrary $\delta_\eta(\omega)$:
	\begin{equation}
	{\mathrm{Re}}[\sigma(0)]_{\text{Drude}\, \eta}=\int \frac{A}{1+B^2 \omega^2}\delta_{\eta}(\omega)d\omega\,.
	\label{prediction}
	\end{equation}
	We can now compare the prediction to our numerical results by first finding the best-fit parameters for $A$ and $B$, and then analyzing how well the Drude model describes the data and whether other theories can describe the data better. We call the fitting procedure a \textit{Drude-based extrapolation} because the estimate of the DC conductivity $\sigma_{\text{DC}}=A$ is the result of the extrapolation of ${\text{Re}}[\sigma(0)]_{\eta}$ from finite $\eta$ to zero. 
	
	We performed the extrapolation for the example described on Fig. \ref{fig:init}. Predicted DC conductivity as a function of the fitting region for several types of broadening $\delta_{\eta}(\omega)$ (see Table \ref{tab:1}) is presented in Fig. \ref{fig:method}(a). The plot contains a region of plateau, where predicted DC conductivity is almost independent of the type of broadening and the extrapolation region. In the calculation, we assumed that the thermodynamic limit for ${\text{Re}}[\sigma(0)]_{\eta}$ is already reached for each $\eta$. This condition seems to be violated for small $\eta$, where predicted conductivity suddenly drops. Certain physical arguments (see Appendix \ref{app:mt}) suggest that the thermodynamic limit of ${\text{Re}}[\sigma(0)]_{\eta}$ is practically reached if the corresponding excitation time of the system is below 
	\begin{equation}
	t_{\text{crit}}= \frac{R}{2v_{\max}}\,,
	\label{t_crit}
	\end{equation}
	where $v_{\max}$ is the maximum group velocity of the band in the direction of the electric field, $R=(L-\lambda_{\beta})/2$ is the effective radius of the system. Here $L$ is the linear size of the system and $\lambda_{\beta}\approx \sqrt{|t|\beta/2}$ is a diffusion length, defined from the evolution of a one-particle equation $\dot{\psi}=-\hat{H}\psi$ at a time $t=\beta/2$ (see Appendix \ref{app:mt} for details). The critical time explains why for the Gaussian and Lorentzian broadening the plateaus are somewhat less \emph{flat} compared to using $\mathrm{sinc}$. The former two functions always have tails in the time domain $t>t_{\text{crit}}$ (see Table \ref{tab:1}), while the sinc function is exactly zero for $|t|>2\eta^{-1}$. Note that the relevant values of $\eta$ in our method are much larger than $\Delta E$. Also, the values of the DC conductivity that our method predicts are almost an order of magnitude larger compared to the classical approach, and, as will become clear soon, much closer to the thermodynamic limit. 
	
	\begin{table}
		\caption{\label{tab:fittingFunctions}%
			Predicted DC conductivity as a function of the broadening function $\delta_\eta(\omega)$ assuming Drude equation for conductivity (\ref{Drude}). 
		}
		
		\begin{tabular}{cccc}
			\hline
			Function name & $\delta_\eta(\omega)\eta$ & $\delta_\eta(t)$ & Predicted $\sigma_{DC}(\eta)$ \\
			\hline
			
			Sinc & $\frac{\sin(2\omega/\eta)}{2\pi\omega/\eta}$ & $\theta(2\eta^{-1}-|t|)$& $A(1-e^{2/B\eta})$\\
			Lorentzian& $\frac{1}{\pi(\omega^2/\eta^2+1)}$&$e^{-\eta |t|}$&$\frac{A}{1+B\eta}$\\ 
			Gaussian& $\frac{1}{\sqrt{\pi}}e^{-\omega^2/\eta^2}$& $e^{-\frac14 x^2\eta^2}$ &$A\frac{\sqrt{\pi}e^{\frac{1}{B^2\eta^2}}\text{erfc}\frac{1}{B\eta}}{B\eta}$\\
			\hline
		\end{tabular}
		
		\label{tab:1}
	\end{table}
	
	The consistency of the Drude-based extrapolation can be confirmed further by studying the convergence of the predicted conductivity with the increase of the size of the system. Fig. \ref{fig:method}(b) shows DC conductivity as a function of the chemical potential for different sizes of the system. For the considered example, the Drude-based extrapolation converges to the thermodynamic limit within 5$\%$ for a cube as small as $6\times 6 \times 6$. The strong oscillations of conductivity for the system of the size $4\times 4\times 4$ can be suppressed by increasing the temperature. 
	
	Let us emphasize that the quality of the extrapolation depends strongly on the chosen model. Using a Gaussian model for conductivity, or simply extrapolating linearly does not produce a plateau and results in predictions highly dependent on the extrapolation region (see Fig. \ref{fig:method}(c)).

	\subsection{Averaging over disorder}
	
	\begin{figure*}
		
		\numberedgraphics{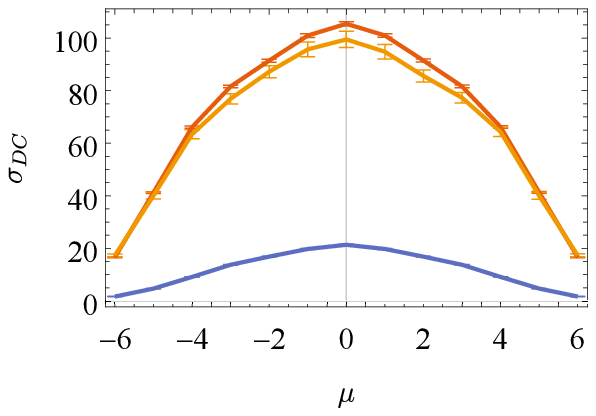}{0.3}{0.24}{-0.08}{a}
		\numberedgraphics{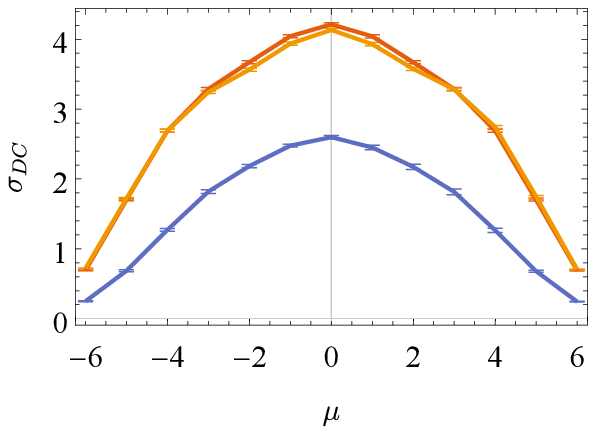}{0.30}{0.2}{-0.08}{b}
		\numberedgraphics{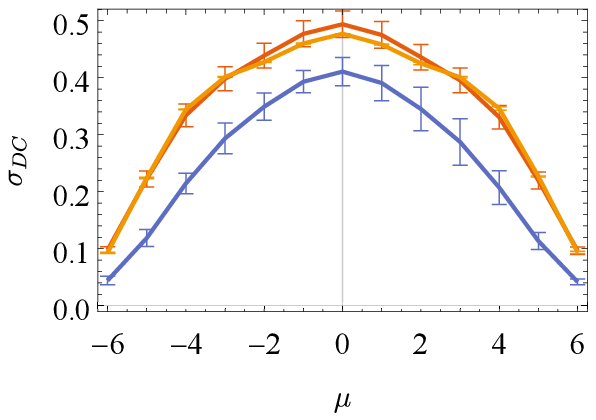}{0.3}{0.24}{-0.08}{c}
		\caption{Conductivity of the system described on the Fig. 1, obtained by three different methods for three values of the onsite disorder: 0.1, 0.5 and 1.5. (a, b and c pictures correspondingly) for a cubic system of the size $10\times10\times10$. The red curve corresponds to our method with Sinc broadening with extrapolation from the region $\eta\in [4v_{max}/L,5v_{max}/L]$, the orange curve - to Landauer based approach, with conductivity defined as $\sigma_{DC}=S\frac{L_1-L_2}{R_1-R_2}$ (see Eq.~(\ref{landauer})), where we took $L_1=250$ and $L_2=150$ for the disorder 0.1, $L_1=14$ and $L_2=20$ for higher disorders and the cross-section $10\times10$. The blue curve - the $\text{Re}[\sigma(\omega)]_{\eta}$ for the Sinc broadening at $\eta=W/L^3$, where $W=12$ is the width of the band. Calculations by the Landauer approach were performed in the Python package Kwant.}
		\label{fig:comparison}
	\end{figure*}
	
	A complementary perspective on the Drude-based extrapolation can be gained through a general connection of conductivity averaged over disorder to the pseudo-conductivity $\text{Re}[\sigma(\omega)]_{\eta}$.
	
	In a mathematical sense, conductivity of a single finite system according to the Kubo-Greenwood Eq.~\eqref{kubo} is not a function, but a functional. Conductivity can be defined as a proper function if averaged over disorder.  Indeed, suppose two energy states $m$ and $n$ have a probability distribution $P(\mathbf{\Gamma}, \mathbf{\Delta})$, where $\mathbf{\Gamma}\equiv\Gamma_{mn}$ and $\mathbf{\Delta}\equiv\Delta_{mn}$ from Eq.~(1). Then we can define averaged conductivity as
	\begin{equation}
	\<\sigma\>(\omega)\equiv\int P(\mathbf{\Gamma}, \mathbf{\Delta})\text{Re}[\sigma(\omega)]d\mathbf{\Gamma}d\mathbf{\Delta}
	\end{equation}
	Using Eq.~(1) and Performing integration over $\mathbf{\Delta}$ results in 
	\begin{equation}
	\<\sigma\>(\omega)=\int P(\mathbf{\Gamma}, \omega)\mathbf{\Gamma}d\mathbf{\Gamma}.
	\end{equation}
	So, if $P(\mathbf{\Gamma}, \omega)$ is a smooth function of $\omega$, then $\<\text{Re}[\sigma(\omega)]\>$ is a smooth function as well. 
	
	Suppose our goal is to find $\<\sigma\>(\omega)$ from sample systems. One of the ways to do so is to find $\<\sigma_{\eta}\>(\omega)\equiv\<\text{Re}[\sigma_\eta(\omega)]\>$ first, and then extrapolate it to the limit $\eta\to 0$. Such extrapolation may benefit from the fact that $\<\sigma_{\eta}\>(\omega)$ is a convolution $\<\sigma\>(\omega)$ with $\delta_\eta(\omega)$:
	\begin{equation}
	\<\sigma_{\eta}\>(\omega)=(\<\sigma\>*\delta_\eta)(\omega),
	\label{conv1}
	\end{equation}
	This can be easily seen from Eq.~(1), and relation
	\begin{equation}
	\delta_\eta(\Delta E-\omega)=(\delta*\delta_\eta)(\Delta E-\omega)
	\end{equation}
	Similarly, averaged conductivity at finite temperature as a function of chemical potential $\mu$ can be expressed as convolution of averaged conductivity at zero temperature with the derivative of Fermi distribution with respect to $\beta$:
	\begin{equation}
	\<\sigma_{\eta \beta}\>(\mu)=(\<\sigma_{\eta \infty}\>*f_\beta') (\mu).
	\label{conv2}
	\end{equation}
	Here $f_\beta'(\mu)\equiv \beta\cosh(\beta \mu/2)^{-2}$ is the derivative of the Fermi distribution over $\beta$. The relation can be obtained by manipulating equation (\ref{kubo_delta_1}).
	Joining Eq.~(\ref{conv1}) and (\ref{conv2}) results in the expression for conductivity with finite broadening and temperature expressed through convolution of averaged conductivity at zero temperature and broadening with the function $f_\beta'(\mu)\delta_\eta(\omega)$: 
	\begin{equation}
	\<\sigma_{\eta \beta}\>(\mu, \omega)=(\<\sigma_{\infty}\>*(f_\beta'\delta_\eta))(\mu, \omega).
	\label{conv12}
	\end{equation}
	Notice, that Eq.~(\ref{conv12}) is valid for any system.
	
	Convolution in the real space corresponds to multiplication in the Fourier space, and therefore convolution suppresses high-frequency harmonics. When we assume a model for $\<\sigma_{\infty}\>(\mu, \omega)$ and then fit the parameters of the model, we effectively fit the low-frequency region in the Fourier space of the model to the low-frequency region of the numerical data. The fitting region has an effective size $\eta^{-1}\times\beta^{-1}$ with the error between the model and the data weighted proportionally to the Fourier transform of $f_\beta'(\mu)\delta_\eta(\omega)$. 
	
	\section{Method comparison}
	We compare our method to two other possible approaches:
	\begin{enumerate}
		\item The first approach follows the logic described in the introduction: we first find $\text{Re}[\sigma(\omega)]_{\eta}$ for several values of $\omega$ at $\eta=W/N$, where $W$ is the band width, and $N$ is the number of energy levels. We then fit the results with a Drude equation (\ref{Drude}) to obtain DC conductivity $\sigma_{\text{DC}}=A$. We already know that conductivity obtained by this method is sensitive to the representation of the delta function and considered $\eta$, but we wanted to explore how much the method underestimates conductivity for different disorders. 
		\item The second approach is based on the Landauer equation \cite{imry1}. Here one finds DC conductance of the system placed between infinite electric leads. Empirically, the resistance of the system can be thought of as a sum of resistances of the bulk and contacts:
		\begin{equation}
		R=R_{\text{bulk}}+R_{\text{contacts}}
		\end{equation}
		If the size of the system is much larger than the correlation length, contact resistance can be assumed constant, while bulk resistance will increase linearly with the length of the system. The conductivity of the system can then be defined as 
		\begin{equation}
		\sigma_{L}\equiv S \left( \frac{d R_{\text{bulk}}}{d L} \right)^{-1}=S \left(\frac{d R}{d L}\right)^{-1}\,.
		\label{landauer}
		\end{equation}
	\end{enumerate}
	We made a comparison of conductivity as a function of chemical potential at three different values of the onsite disorder obtained by different methods. The results are presented in Fig. \ref{fig:comparison}. The values of conductivity obtained by our method and the Landauer-based approach are within the error-bars, while the classical approach underestimates conductivity by a factor of 5 for the weak disorder, and becomes closer to other methods for high disorders. 
	
	\begin{figure}
		\numberedgraphics{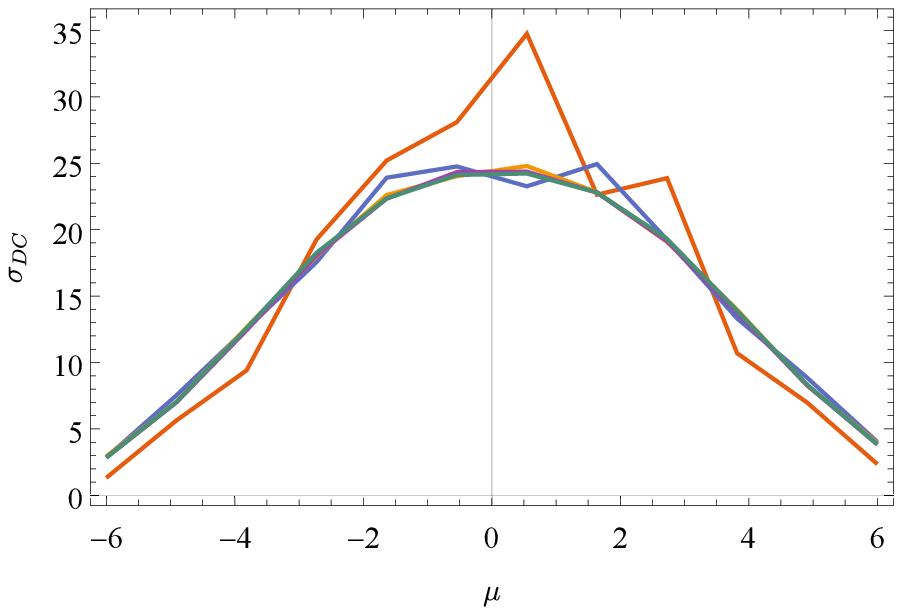}{0.45}{0.18}{-0.08}{a}
		\numberedgraphics{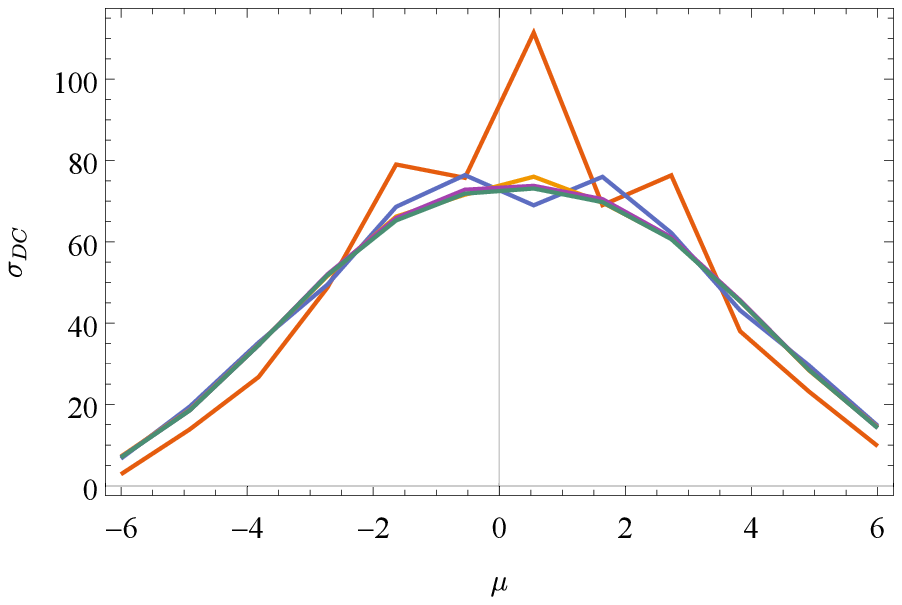}{0.45}{0.18}{-0.08}{b}
		
		\caption{Conductivity as a function of the chemical potential of a tight binding model with strong rare impurities for different sizes of the system. The system is identical to the one described in Fig. 1 with the exception that onsite potential has a probability $p$ to be equal to $V(1-p)$ and a probability $1-p$ to be equal to $-Vp$. We took $V=100$ and $p=0.003$ for plot (a) and $p=0.001$ for plot (b). The red curve corresponds to a system of size $4\times 4 \times 4$, blue - $6\times 6 \times 6$ and so on. Comparing these results to Fig. \ref{fig:method}(b) it is clear that appearance of a new length scale does not have any impact on the convergence rate of the method.}
		\label{fig:comparison1}
	\end{figure}

	\begin{figure}
		
		\numberedgraphics{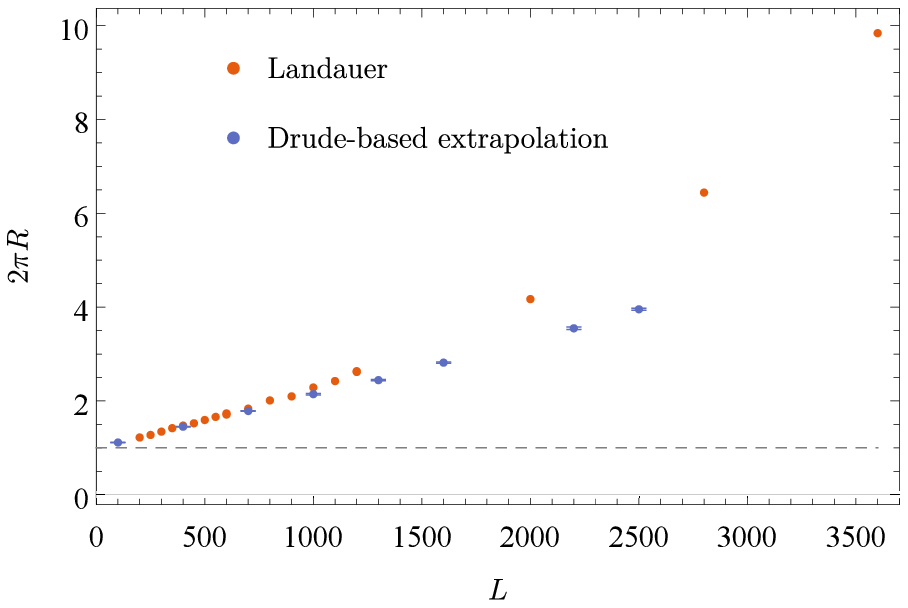}{0.45}{0.18}{-0.08}{a}
		\numberedgraphics{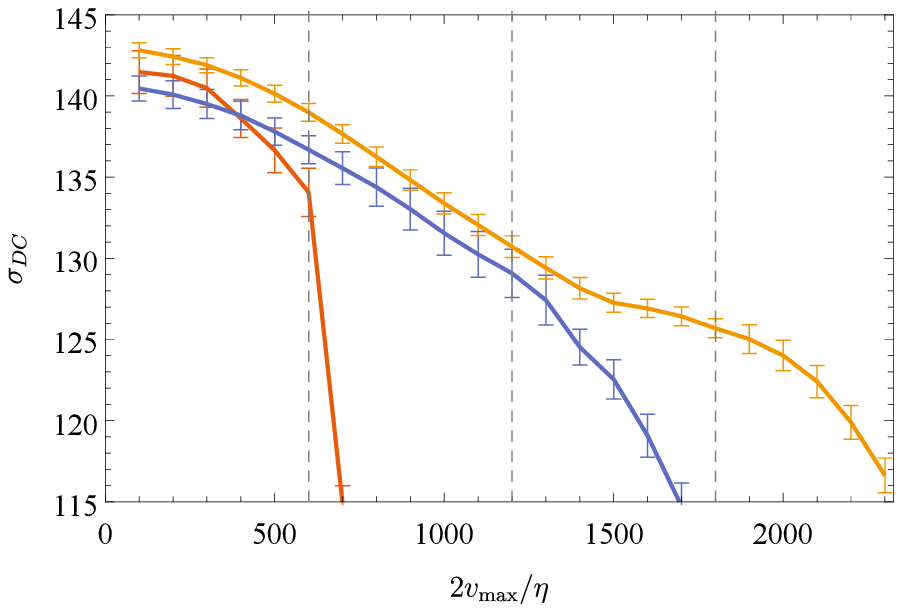}{0.45}{0.18}{-0.16}{b}
		
		\caption{a) Resistance as a function of the length of 1D chain obtained by Landauer-based approach (red dots) and Drude-based extrapolation (blue dots). The system is a 1D tight binding model with neared neighbor hopping $t=-1$, chemical potential $\mu=-1$ and onsite disorder uniformly distributed in the range $[-\Delta \varepsilon,\Delta \varepsilon]$, where $\Delta \varepsilon=0.1$. Drude based extrapolation was performed from the points $\eta=10/L$ and $\eta=15/L$. Resistance was calculated by the equation $R=(2\pi)^{-1}+L/\sigma_{\text{DC}}$. b) Drude-based conductivity of the 1D chain as a function of the delta function broadening $\eta$ for the Sinc broadening (see Table \ref{tab:1}). The horizontal axis corresponds to $v_{\max} \tau$, where $\tau=2/\eta$ is the time over which the electric field was on before current measurement, and $v_{\text{max}}=2|t|$ is  the maximum group velocity in the band. Extrapolation performed between $\eta$ and the neighbor point $\eta+\Delta\eta$, $\Delta\eta>0$. Different curves correspond to different length of the chain: red curve - $L=600$, blue - $L=1200$, orange - $L=1800$.  }
		\label{fig:comparison2}
	\end{figure}
	
	\section{Applications}
	We applied our method to two test systems. The first is a 3D system with a low concentration of impurities, and the second is a 1D disordered chain. The essential features of the models are the appearance of new length scales: the typical distance between impurities and localization length.
	
	\subsection{3D system with impurities}
	The model of the uncorrelated weak onsite disorder, which we used in the previous examples, is not very realistic for real materials. Another common approach is to model disorder by a small concentration of strong impurities in an otherwise perfect system. We were interested whether the addition of the new length scale would influence the convergence rate of the method. Intuitively it seems impossible that the conductivity of a system smaller than an average distance between impurities can be close to the thermodynamic limit. On the other hand, the intuition about the convergence of $\text{Re}[\sigma(\omega)]_{\eta}$ described in Appendix B did not rely on any specific properties of the disorder distribution.  
	
	We considered a tight binding 3D system of the size $L\times L\times L$ and nearest neighbor hopping similar to the system described in Fig. 1, but with the difference that onsite potential could be either $V(1-p)$ with a small probability $p$ or $-Vp$ with probability $1-p$. With such choice the average potential is zero, while the difference between "impurities" and background potential is $V$. We considered two examples with $p=0.003$ and $p=0.001$. Calculated conductivity as a function of chemical potential for different sizes of the system is presented on Fig. \ref{fig:comparison1}. Comparing the results to Fig. \ref{fig:method}(b) it is evident that convergence rate is not influenced by the introduction of a new length scale. 
	
	\subsection{1D system with disorder}
	Often an intuition about a 3D or 2D system can be obtained by looking at a 1D system with a similar structure. Unfortunately, 1D systems with disorder have zero DC conductivity in thermodynamic limit due to the localization of the wave-functions \cite{ande1}. On the other hand, short chains have an empirically well defined conductivity, since their resistance increases linearly with the size of the system. This result is demonstrated on Fig. \ref{fig:comparison2}(a)), where we plotted Landauer resistance vs length of a 1D tight binding chain with nearest neighbor hopping $t=-1$, onsite disorder $\Delta \varepsilon=0.1$, chemical potential $\mu=-1$ and temperature corresponding to $\beta=30$.
	
	We were interested to what extend the Drude-based extrapolation would reproduce these results. It appears conductivity obtained by our method predicts resistance of short chains very well (see blue dots on Fig. \ref{fig:comparison2}(a)), but underestimates conductivity for long chains. This is expected, since long chains are insulators and conductivity cannot follow the Drude equation for long excitation times. This becomes also evident from the plot of conductivity vs fitting region (Fig. \ref{fig:comparison2}(b)): while in 3D systems conductivity was almost independent of the fitting region, in 1D conductivity predicted from long and short excitation times is different.
	
	The important result is that we can predict static response of systems connected to the leads from dynamic response of the system with periodic boundary conditions. This is not the case with the classical approach, which noticeably underestimates the DC response.
	
	\section{Summary and outlook}
	
	In the current paper, we proposed a procedure for estimating the thermodynamic limit of DC conductivity from a dynamic current response of a microscopic quantum system. The following steps roughly describe the procedure: 
	\begin{enumerate}
		\item First, conductivity $\sigma_{\mathrm{DC}\eta}$ of the microscopic system is found using Kubo-Greenwood equation \eqref{kubo_delta0} with smeared delta-function $\delta(\omega)\rightarrow \delta_{\eta}(\omega)$ for several values of the broadening parameter $\eta$. The conductivity $\sigma_{\mathrm{DC}\eta}$ is then averaged over many realizations of the disorder.
		
		\item A particular model for conductivity $\sigma(\omega)$ is adopted (Drude model in our case), and parameters of the model (and thus DC conductivity) are found by fitting $\sigma_{\mathrm{DC}\eta}$ calculated by Eq. \eqref{conv1} to numerical results. 
	\end{enumerate}
	Consistency of the procedure can be verified by considering the convergence of the obtained DC conductivity as the size of the system increases similarly as in the classical approach. 
	Importantly, the convergence is accelerated greatly if the broadening parameter $\eta$ and the temperature are above a certain threshold. Physical considerations (see \ref{app:mt}) suggest that the critical value of $\eta$ is such that the Fourier transform of $\delta_\eta(\omega)$ -- $\delta_\eta(t)$, has a width of $L/v_{\mathrm{max}}$, where $L$ is the size of the system and $v_\mathrm{max}$ - maximum group velocity in the corresponding direction. The critical temperature is $T_{\mathrm{crit}}=2D/L^2$, where $D$ is the effective diffusion coefficient of the Schrodinger equation in the imaginary time. 
	
	Another major factor influencing the convergence rate is the chosen model for conductivity and the type of the broadening function. In the considered example of a 3D tight binding model, the best performers were the Drude model for conductivity and the sinc broadening. In real systems, the situation may change due to the presence of optical resonances. Gaussian broadening, because of being more localized, may perform better, making the final result less sensitive to a wrong choice of conductivity model. The question, though, requires further investigation.   
	
	So far, we considered Drude-based extrapolation in the context of the thermodynamic limit. Another way of looking at the extrapolation is through a lens of data compression: Drude model predicts an electric current response of a finite system to a family of slowly varying electric fields acting over the times $t<t_{\text{crit}}=L/2v_{\text{max}}$, thus compressing the information about the responsiveness of the system to two numbers. Notably, the value of conductivity obtained by our method also matches the Landauer-based approach, which implies a static response of a system connected to electrical leads can be predicted from the dynamic response of the system detached from the leads. It is not the case with a classical approach, which predicts a much lower value for DC conductivity than Landauer.
	
	While the Drude-based extrapolation works well in practice, a rigorous understanding is still lacking. The arguments presented in \ref{app:mt} provide a strong motivation for the convergence conditions of our method to thermodynamic limit, but we do not have mathematical proof. Looking at the plateau in our plots, one may wonder if there is a way to define DC conductivity unambiguously, even for finite systems. Unfortunately, we do not see any physical basis for this claim, since the Kubo-Greenwood equation is exact. Experimental definition of finite-sized conductivity contains similar ambiguities to the theoretical approaches: one needs to exclude contact resistance and interaction of the system with the thermostat. The empirically well-defined plateau only means that $\sigma(\omega)_{\eta}$ is indeed very close to the thermodynamic limit and that our assumptions about the shape of conductivity as a function of frequency are close to the actual shape. 
	
	We believe that an important application of our method is its extension to the Kernel Polynomial based Methods (KPM) \cite{alex1} and other versions of the Kubo-Greenwood formula \cite{diem1}. The KPM methods allow calculation of conductivity of very large systems at low temperatures by removing the necessity of calculating the whole spectrum and scaling linearly with the system size.
	
	The application of our method to the existing approaches in DFT packages \cite{cald1} should be straightforward. Calculations of conductivity at different broadenings of the delta function are performed routinely as a consistency check of the classical approach. Adding the extrapolation step on the top is a computationally efficient way to greatly improve convergence to the thermodynamic limit.
	
	\section{Acknowledgements}
	
	The author thanks Jorge Sofo for proposing this problem and numerous helpful discussions. We thank Lucas Hackl for the help with structuring and preparing this manuscript. We thank Barbara Kennedy for the help with the language and outlook. We thank Hyokun Yun, William Faugno, Jacob Robins, Brett Green, Likun Zhang, Dmytro Oliinychenko and Xiaoxiao Li for productive discussions, which helped to clarify various aspects of the work.

	\section*{References}
	
	\bibliography{StandardConducitivityModel}
	
	\clearpage
	
	\appendix
	\section{Kubo-Greenwood equation derivation}
	\label{app:kg}
	\subsection{Linear response in quantum systems}
	
	Consider a system with discrete energy spectrum described by a quantum Hamiltonian $\hat{H}_0$, perturbed by a time-dependent term $\varepsilon \hat{V} f(t)$, where $\varepsilon$ is a small parameter, and $f(t)$ some function of time. We want to find a response of an observable $\hat{X}$, specifically we look for  $\<\hat{X}\>(t)\equiv\text{tr}(\hat{\rho}(t) \hat{X})$, where $\hat{\rho}(t)$ is the density matrix of the system. From the interaction picture, one can show the average value of the observable is
	
	\begin{equation}
	\<\hat{X}\>(t)=\<\underrightarrow{ e}^{i\varepsilon\int_{-\infty}^{t} \hat{V}_I(t_1) f(t_1)dt_1}\hat{X}\underleftarrow{ e}^{-i\varepsilon\int_{-\infty}^{t} \hat{V}_I(t_1) f(t_1)dt_1}\>_0.
	\label{pert1}
	\end{equation} 
	Here $\<\hat{A}\>_0\equiv \text{tr}(\hat{\rho}_0 \hat{A})$, where $\hat{\rho}_0=e^{-\beta \hat{H}_0}/Z$ is the equilibrium density matrix of unperturbed system. $\hat{V}_I(t)\equiv e^{i\hat{H}_0t}\hat{V}e^{-i\hat{H}_0t}$ is the operator $\hat{V}$ in interaction representation. Arrows under exponents indicate the direction of the time ordering, for example:
	\begin{equation}
	\underleftarrow{ e}^{\int_{0}^{t} \hat{V}(t_1) dt_1}\equiv\lim\limits_{\delta\to 0} (1+\delta \hat{V}(t-\delta ))...(1+\delta  \hat{V}(\delta ))(1+\delta  \hat{V}(0)).
	\label{lim0}
	\end{equation} 
	This is not a standard notation for time ordering, but we find it more intuitive since the arrow indicates the direction in which time increases in the limit expansion.
	
	Expanding Eq.~(\ref{pert1}) to the first order in $\varepsilon$ and shifting time to $t=0$ yields 
	\begin{equation}
	\<\hat{X}\>=\<\hat{X}\>_0 -i \varepsilon \int_{-\infty}^{0}\<[\hat{X}, \hat{V}_I(t_1)] \>_0 f(t_1)d t_1
	\label{pert2}
	\end{equation}
	where $[\hat{A},\hat{B}]\equiv \hat{A}\hat{B}-\hat{B}\hat{A}$. 
	
	\subsection{\label{sec:spec}Spectral representation}
	Let us find the response of the system at a frequency $\omega$. 
	We allow $f(t)$ to be complex and take $f(t)=e^{(i \omega+\eta) t}$, with $\eta>0$. Taking finite but small $\eta$ corresponds to perturbation turned on slowly and assures convergence of the integral (\ref{pert2}). 
	
	In spectral representation we express Eq.~(\ref{pert2}) in the basis of eigenstates of Hamiltonian $\hat{H}_0$: $\hat{H}_0|n\>=E_n|n\>$. We define $\<n|\hat{X}|m\>\equiv X_{nm}$ and $\<n|\hat{V}|m\>\equiv V_{nm}$, then $\<n|\hat{V}_I|m\>\equiv V_{nm}e^{-i (E_m-E_n) t}$. The density matrix $\<n|\hat{\rho}_0|m\>=\delta_{mn}e^{-\beta E_n}/Z$, where $\delta_{mn}=1$ if $m=n$ and $0$ otherwise. Substituting these quantities into equation (\ref{pert2}), performing integration and relabeling some of the summation indexes yields
	
	\begin{equation}
	\<\hat{X}\>=\<\hat{X}\>_0+\frac1Z\sum_{mn} \frac{e^{-\beta E_n}-e^{-\beta E_m}}{E_n-E_m-\omega + i \eta} X_{nm}V_{mn}
	\label{spec1}
	\end{equation}
	
	\subsection{\label{sec:cond} Conductivity}
	
	Let us apply linear response to conductivity. We are interested in a tight binding model with spinless electrons with random hoppings and site energies and periodic boundary conditions. No interactions or averaging over disorder is assumed at this point. The derivation is not limited to such systems, but it is easier to have something specific in mind. 
	
	Electric field is created by a change of vector potential $\vec{E}=-d\vec{A}/dt$.
	\subsubsection{Current operator}
	We postulate the Hamiltonian $\hat{H}(A)$ to depend on vector potential by Peierls substitution, i.e. each hopping coefficient $t$ is a function of $A$: 
	\begin{equation}
	t(\vec{A})=te^{i(\vec{A},\vec{a})}
	\end{equation}
	where $(\vec{A},\vec{a})$ is a scalar product of the vector potential $\vec{A}$ and a vector $\vec{a}$ connecting the corresponding cites or atoms.
	
	Without loss of generality, we can assume the electric field to be in the $x$ direction. For the scope of this paper, we are interested only in the current in the $x$ direction. The corresponding averaged current density operator is
	\begin{equation}
	\hat{j}=-\frac{1}{V}\partial_A \hat{H},
	\label{j_def}
	\end{equation} 
	where V is the volume of the system. The choice of the current density operator ensures that energy production rate in the system is equal to $d\<H\>/dt=\<\hat{j}\>E V$. The last equation can be directrly checked by the substitution of Eq.~\eqref{j_def}.
	
	\subsubsection{Linear response of the current}
	
	Assuming $A$ to be small we can Taylor expand $\hat{H}$ and $\hat{j}$ at $A=0$: 
	\begin{eqnarray}
	\hat{H}=\hat{H}_0-AV\hat{j}_0,\\
	\hat{j}=\hat{j}_0-A\hat{M}
	\end{eqnarray}
	where $\hat{M}\equiv V^{-1} \partial_A^2 \hat{H}|_{A=0}$. We see that the perturbation in our case is $-AV\hat{j}_0$. Reconciling this with equation (\ref{pert2}) to the first order in $A$ yields
	\begin{equation}
	\<\hat{j}\>=-\<\hat{M}\>_0 A(t)+iV \int_{-\infty}^{0}\<[\hat{j}_0,\hat{j}_{0I}(t_1)]\>_0 A(t_1)dt_1.
	\label{DC1}
	\end{equation}
	
	The first term is usually called \textit{diamagnetic current} and the second term is \textit{paramagnetic current}.
	
	\subsubsection{AC response}
	
	Let us find the response to the vector potential $A(t)=-e^{\eta t +i\omega t} E_0/(\eta +i\omega) $. After going
	to spectral representation in Eq.~(\ref{DC1}), performing integration and rearranging some terms we obtain the following result for the current:
	\begin{equation}
	\frac{\<\hat{j}\>}{E_0}= O - \frac{V}{Z} \sum_{nm} \frac{e^{-\beta E_n}-e^{-\beta E_m}}{E_n-E_m}  \frac{i j_{nm}j_{mn}}{E_n-E_m+i\eta - \omega},
	\label{1}
	\end{equation}
	where
	\begin{equation}
	\begin{split}
	&O=\frac{Z^{-1}}{\eta + i\omega}\times \\ &
	\left(\sum_n M_{nn} e^{-\beta E_n}+\sum_{nm} \frac{e^{-\beta E_n}-e^{-\beta E_m}}{E_n-E_m}j_{nm}j_{mn}\right).
	\end{split}
	\end{equation}
	The sum in the brackets can be shown to be $\partial\<j\>_0/\partial A$. Usually, equilibrium current is assumed to be zero for any $A$. This is not necessarily true for a system with periodic boundary conditions, and there are current oscillations periodic in magnetic flux quanta through the system. The second term in Eq.~(\ref{1}) also oscillates with the number of flux quanta through the system. Numerically, the oscillations and the term $O$ die out at sufficiently high temperatures. Also, the term can be shown to be exactly zero, if averaged over vector potential. We will ignore these effects for now as they are out of the scope of this paper.
	
	The final expression for the current is the standard Kubo-Greenwood formula:
	\begin{equation}
	\frac{\<\hat{j}\>}{E_0}=-\frac VZ\sum_{nm} \frac{e^{-\beta E_n}-e^{-\beta E_m}}{E_n-E_m}\frac{i j_{nm}j_{mn}}{E_n-E_m+i\eta-\omega}.
	\label{kubo3}
	\end{equation}
	
	For a system of non-interacting electrons equation (\ref{kubo3}) reduces to 
	\begin{equation}
	\frac{\<\hat{j}\>}{E_0}=-\sum_{nm} \frac{f_n-f_m}{\varepsilon_n-\varepsilon_m}\frac{i j_{nm}j_{mn}}{\varepsilon_n-\varepsilon_m+i\eta-\omega}.
	\label{kubo4}
	\end{equation}
	where $\varepsilon_n$ are one-electron energies, and $f_n=(1+e^{\beta (\varepsilon_n-\mu)})^{-1}$ - Fermi distribution. The real part of the current is 
	\begin{equation}
	\text{Re}\left[\frac{\<\hat{j}\>}{E_0}\right]=-\sum_{nm} \frac{f_n-f_m}{\varepsilon_n-\varepsilon_m}j_{nm}j_{mn}\frac{\eta }{(\varepsilon_n-\varepsilon_m-\omega)^2 + \eta^2}.
	\label{kubo5}
	\end{equation}
	
	In the limit $\eta\to 0$ this yields: 
	\begin{equation}
	\text{Re}\left[\sigma(\omega)\right]=-\pi \sum_{nm} \frac{f_n-f_m}{\varepsilon_n-\varepsilon_m}j_{nm}j_{mn}\delta(\varepsilon_n-\varepsilon_m-\omega),
	\label{kubo_delta}
	\end{equation}
	which corresponds to equation (\ref{kubo}).
	
	\section{The critical excitation time for Drude-based extrapolation}
	\label{app:mt}
	
	In the core of our method lies an assumption that if the delta function broadening is sufficiently large, the pseudo-conductivity $\text{Re}[\sigma_{\text{DC}}(0)]\eta$ is practically converged to the thermodynamic limit. This appendix finds an estimate for the critical broadening $\eta$, specifically, we show that if the system is excited over a finite period of time $t<t_{crit}\approx L/2v_{max}$ the electric current response must be practically independent of $L$ if the temperature is high enough.
	
	Let us start with equation (\ref{DC1}). We can break operators $\hat{j}$ and $\hat{M}$ into a sum of local operators: 
	\begin{equation}
	\begin{split}
	\hat{j}=\frac1V\sum_{\vec{r}}\hat{j}_{\vec{r}},\\
	\hat{M}=\frac1V\sum_{\vec{r}}\hat{M}_{\vec{r}},
	\end{split}
	\end{equation}
	where the sum is taken over all unit cells. The breaking can always be performed in many different ways. It is only important that the local operators spread over the finite number of cells.
	From here we can plug these sums into the equation (\ref{DC1}) and average the equation over disorder. If the distribution of the disorder is translationally symmetric, the final sum will consist of identical terms, each of which can be equalized individually:
	\begin{equation}
	\begin{split}
	\<\hat{j}_{\vec{r}}\>=&-\<\hat{M}_{\vec{r}}\>_0 A(t)+\\&
	i \sum_{\Delta\vec{r}}\int_{-\infty}^{0}\<[\hat{j}_{0\vec{r}+\Delta \vec{r}},\hat{j}_{0I\vec{r}}(t_1)]\>_0 A(t_1)dt_1.
	\end{split}
	\label{kubo_local}
	\end{equation}
	Notice, that the equation expresses the current response at a single site through the correlator of the current operator at the same site propagated in time with current operators at neighbor sites. Sometimes, the equation can be simplified even further if the system possesses a symmetry. For instance in the case of cubic symmetry it is enough to know the current terms only for $\Delta \vec{r}$ in the positive $x$, $y$ and $z$ directions. We are considering here a more general, non-symmetric case.
	
	Next, our intuition goes the following  way.
	Lets assume the system consists of non-interacting particles and there is no disorder. In the expression \eqref{kubo_local} we need to take a trace of the operator (see the second term):
	\begin{equation}
	J_{\vec{r}\beta}=e^{-\frac{\beta H}{2}}J_{\vec{r}}e^{-\frac{\beta H}{2}}\,,
	\label{J_rbeta}
	\end{equation}
	where 
	\begin{equation}\begin{split}
	J_{\vec{r}}\equiv\sum_{\Delta\vec{r}}[\hat{j}_{0\vec{r}+\Delta \vec{r}},e^{i\hat{H}t_1}\hat{j}_{0I\vec{r}}e^{-i\hat{H}t_1}].
	\end{split}
	\end{equation}
	Consider the operator $\hat{U}=e^{i\hat{H}t}$ acting on a single particle state $|\vec{r}\>$, localized at the site $\vec{r}$. The operator propagates the state in all directions at a speed of $v_{\max}$ for a time $t$, and therefore spreads the state over the sites in a sphere of a radius $v_{\max}t$. Here $v_{max}$ is the maximum group velocity of the band. From the properties of one-particle operators in non-interacting systems it follows that the operator $J_{\vec{r}}$ is localized in the same sphere, meaning it  consists of creation and annihilation operators that are located inside the sphere. 
	
	In order to find the current response we need to calculate the trace:
	\begin{equation}
	\begin{split}
	\text{tr}(J_{\vec{r}\beta}) &\equiv  \sum_{i}\<i|e^{-\frac{\beta H}{2}}J_{\vec{r}}e^{-\frac{\beta H}{2}}|i\>\\&
	=\sum_{i}\<i_{\beta}|J_{\vec{r}}|i_{\beta}\>\,.
	\end{split}
	\end{equation}
	Let us understand the structure of states $|i_{\beta}\>=e^{-\beta \hat{H}/2|i\>}$: if a state $|i\>$ contains particles at certain positions, then the operator $e^{-\beta \hat{H}/2}$ will "diffuse" the particles for a time $\beta/2$ with diffusion coefficient $D$ proportional to the hopping matrix elements. Since the operator $J_{\vec{r}}$ is localized in a sphere of a radius $v_{\max}t$, a single state average $\<i_{\beta}|J_{\vec{r}}|i_{\beta}\>$ should be practically independent of the distribution of particles outside the critical radius $R_{\text{crit}}=v_{\max}t+\sqrt{D\beta/2}$, thus making the trace independent of the size of the system if it is above the critical radius.
	
	Similar logic can be applied to the operator $M_{\vec{r}}$, which concludes our argument that $\<\hat{j}_{\vec{r}}\>$ in Eq.~(\eqref{kubo_local}) should be practically independent of the size of the system for $L\gtrsim 2R$ in the case when the disorder is absent. Introduction of disorder can only hinder the excitation propagation and diffusion, and therefore we expect the convergence to be even faster for disordered systems. Notice that numerically, our method usually converged even better, with conductivity independent of the size of the system for $L\ge R$ rather than $L\ge 2R$ (see Fig. \ref{fig:method}(a) and \ref{fig:comparison2}(b)). We believe this is associated with the aforementioned symmetries of the system. 
	
\end{document}